\documentstyle[times,epsf]{l-aa}

\begin{document}

  \thesaurus{01	          
	        (13.25.5; 
		08.09.2;  
 		08.02.1;  
		02.01.2)} 
   \title{Attenuation of dipping at low energies in the LMXB source
          \hbox {X 1755-338}}

   \subtitle{}

   \author{M. J. Church \and M. Ba\l uci\'nska-Church}

   \offprints{M. J. Church}

   \institute{School of Physics and Space Research,
              University of Birmingham,
              Edgbaston, Birmingham B15 2TT, UK}

   \date{Received ; accepted }

   \headnote{{\it Letter to the Editor}}

   \maketitle

   \begin{abstract}
We report spectral fitting results for {\it Rosat PSPC} observations of the 
dipping source X 1755-338. These results are consistent with the two-component
model that we previously proposed consisting of a blackbody point source 
plus an extended power law component. Remarkably, the low energy cut-off of 
the spectrum does not change appreciably in dips, and it can be seen that
dipping takes place in the higher part of the PSPC band where the blackbody
contributes to the spectrum. Thus dipping consists primarily of
absorption of this component, whereas the low energy cut-off is determined
by the power law. Thus, above 0.5 keV, the dipping is approximately independent 
of energy as seen in {\it Exosat} and $\rm {10\pm 0.4}$\% deep. Below 0.5 keV, 
there may be a small residual dipping effect of up to 3\%; however dipping is
certainly substantially reduced compared with the 20\% seen by {\it
Exosat},
and this is the first time that the
effective cessation of dipping at low energies in such sources has been
seen.


   \end{abstract}

\section{Introduction}

X 1755-338 is one of the most interesting members of the class
of $\sim $ 10 Low Mass X-ray Binary (LMXB) sources which show periodic dips
in X-ray intensity at the orbital period. It is generally accepted that
the dips are due to absorption in the bulge in the outer accretion disc
caused by the impact of the accretion flow from the companion (White \&
Swank 1982). The evolution of the X-ray spectra during dips varies
considerably between dipping sources and it is not the case that the
sources in general show an increase in low energy absorption leading to a 
hardening of the spectrum as expected for photoelectric absorption in the
relatively cool material of the outer disc. In particular cases there may
be a hardening or a softening or no change in hardness. X 1755-338
is very unusual in this respect since the high quality {\it Exosat ME} spectra
showed that within error, the dipping was energy independent in the band 1
- 10 keV (White et al. 1984), because of which it has been called the
energy-independent dipper. Various explanations of this effect have been
proposed, notably that the metallicity of the absorber is substantially
reduced from solar values (White et al. 1984). Other possibilities include absorption in a region closer to the
central compact object where there will be stronger photoionization
(Frank, King \& Lasota 1987), or partial covering of an extended source
(Frank \& Sztajno 1984). More recently we have suggested an explanation
based on a two-component model of the source (Church \& Ba\l
uci\'nska-Church 1993) in which the total emission consists of a blackbody
point source originating in the neighbourhood of the compact object plus an
extended power law component probably due to Comptonisation in an Accretion
Disc Corona. Dipping is seen to be primarily due to absorption of the blackbody.

\begin{figure*}
\epsffile{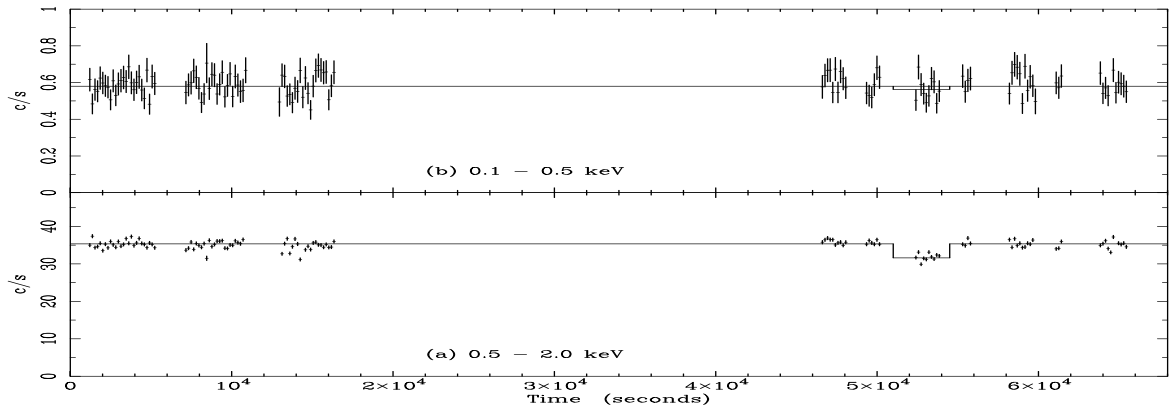}
\caption[]{X-ray light curves for the complete 18 h {\it Rosat} observation in
two energy bands: 0.5 - 2.0 keV and 0.1 - 0.5 keV with 160 s timebins.}
\end{figure*} 

\par We have found that the same two-component model can also explain the 
dipping in X 1624-490 in which the spectral evolution is complex, consisting
of a hardening in shallow dipping, followed by a softening in deeper dipping
(Church \& Ba\l uci\'nska-Church 1995a). It can also explain dipping in the
very different source XB\thinspace 1916-053 in which dipping can reach
100\% (Church et al. 1997), and we have proposed that the two-component model 
may explain all of the dipping sources (Church \& Ba\l uci\'nska-Church 1995b).
It is important to determine which explanation of the energy-independence 
is correct, and one way to do this is to examine the spectrum at energies
both lower and higher than the {\it Exosat ME} band. If the two-component model
is correct, at photon energies much lower and higher 
than $\rm {kT_{bb}}$ for the
blackbody (0.9 keV) there should be very little dipping since the blackbody
contribution to the spectrum will be negligible.

\par In the present paper we examine the spectrum of the source in the
PSPC band 0.1 - 2.0 keV. We find that the two-component model is a good
representation of the spectra. Dipping is less obvious in the PSPC band
than in the ME, but it can be seen that it is due to increases of $\rm {N_H}$
for the blackbody, and that at the lowest energies below 0.5 keV,
the extent of dipping is substantially reduced.

\section{Data analysis}
Three observations have been made of X 1755-338 with the
{\it Rosat PSPC}. In two of these there is little or no sign of dipping. 
We present results for the other longer observation made on 1993 March 29, 
lasting 18 h. Source data were extracted from a circle of radius 2 arcmin, 
taking into account the dust scattered X-ray halo of the source discussed by 
Predehl and Schmitt (1995). Photons scattered in the halo are not expected to 
show dipping because of the variable time delays introduced which smear out 
time-variability. From the figure given by Predehl and Schmitt, it can be
estimated that a radius of 2 arcmin includes 96\% of the unscattered
source counts, but excludes 97\% of the halo, which should contribute about
3 c/s in the total PSPC band. For source extraction from a 2 arcmin region,
background subtraction is not very important in this bright source, 
normally less than 0.1\% of the total count rate, except
for the subtraction of sharply rising background at the ends of each
of the first 3 sections of data in Fig. 1 which in fact caused the
switch-off of the detectors in the following data gaps. We obtained
background data from an annulus between 0.22$^\circ $ and 0.28$^\circ $, 
excluding point sources outside this region, and excluding most of the
halo which Predehl and Schmitt show is becoming small at 0.22$^\circ $.
It is important for spectral fitting that background subtraction is made 
from a region with sufficiently high background count, and this region
is the best for this requirement. 

\par The light curve of the source in the band 0.5 - 2.0 keV is shown in Fig. 1a
with 160 s timebins. Ideally, longer binning would be better, but the
shortness of some of the data sections do not allow this. Consequently, it
is still possible to see the effects in the light curves of the wobbling of the
telescope to prevent occultation of sources behind the wires, and this 
effect dominates over Poisson noise in the light curves.
A section of data can be seen with count rate reduced by $\sim $ 10\% ,
which, allowing for data gaps, lasts
between 25 and 68 m. The duration is consistent with the duration
of dipping seen in the {\it Exosat ME} of 40 m. The depth of dipping in the
ME was typically $\sim $ 20\% (1 - 10 keV), although one dip at 12\% was
seen. Thus there is an indication that the depth of dipping is less in the
0.1 - 2.0 keV band. The lack of other dips is consistent with
the orbital period of about 4.4 h which predicts that other dips would 
occur during the data gaps (although there is a possibility that the
first section of data in the light curve is contaminated at the
end by the onset of dipping). Finally, our spectral fitting results are fully 
consistent with this low intensity period being a dip.

\par We extracted spectra for each of the non-dip sections of data in the 
light curve lasting more than 1500 s, and also dip data (1700 s). 
These were corrected by subtracting the background and correcting for 
deadtime and vignetting. The spectra were rebinned into 24 channels
(after excluding channels below 0.1 keV and above 2.0 keV), and a 
systematic error of 2\% was added conservatively to each spectral channel.
The results were completely consistent with those obtained
fitting the spectra in primitive channels without rebinning. 
We present results below obtained using the January 1993
instrument response as appropriate to this observation in the later part
of the {\it Rosat} mission (see Fiore et al. 1994). However to assess
systematic errors, the data were also analysed using the alternative
March 1992 response. It was found that parameters derived from spectral
fitting changed by only 1.5\% to 5\%, giving confidence in the results.

\section{Results}

\par Firstly, we consider simple models.
In the case of the PSPC non-dip spectra of X 1755-338,
a simple absorbed power law model gives an acceptable fit, with however 
a value of $\Gamma $ = 1.9$\pm $0.1. In our previous work on the {\it
Exosat ME} data, we found that a two-component model was necessary to
describe the source, consisting of a blackbody with $\rm {kT_{bb}}$ of
0.88 keV, and a power law with photon index $\Gamma $ of 2.67.
Having re-examined the ME data we
find that a value of 1.9 can be rejected for a single spectrum with
certainty greater than 93\%, and in our {\it Exosat} work, we were
able to fit the complete observation divided into a sequence of 314 spectra 
with $\Gamma $ values between 2.6 and 2.8. Thus the simple power law fit 
is not acceptable. It can however be used to test whether the low energy 
cut-off (LECO) of the spectrum changed between non-dip and dip data
(minimising model dependency). There was {\it no detectable change 
in $\rm {N_H}$}, but dipping could be seen as a decrease in the spectral
flux in the higher part of the PSPC band (0.5 - 2 keV).

\par Use of a two-component model is clearly indicated. It is shown below 
(see Fig. 2) that the LECO is determined by the power law, and so the 
blackbody can not have 
a smaller $\rm {N_H}$ than the power law, and  a model of the form 
$\rm {AB\cdot (AB\cdot BB+PL)}$ (where AB, BB and PL are the absorption,
blackbody and power law terms) should be used. Very good fits were obtained 
for both non-dip and dip data using this model, showing that the PSPC data are 
fully consistent with the two-component model. If $\Gamma $ and 
$\rm {kT_{bb}}$ are allowed to be free, low  values of $\Gamma $ are again 
obtained $\sim $ 1.5, inconsistent with the values we obtained from the ME 
data. The {\it Rosat PSPC} is not able to constrain power law index well in 
the presence of a second component, or $\rm {kT_{bb}}$ for a blackbody
peaking above the PSPC band.
However the PSPC data can be used in conjunction with parameter values 
already established for a source to determine the 
lower energy part of the spectrum. It is clearly more sensible to fix 
both $\Gamma $ and $\rm {kT_{bb}}$ at the ME values. 

\par Results are shown in Fig. 2 (plotted with primitive channels, see
below). In Fig. 3, the spectra analysed with 24 channels are shown,
plotted together with typical {\it Exosat ME} spectra.
Again, there was little or no change in the LECO between non-dip
and dip spectra. The energy range of the {\it Rosat} PSPC is ideal for 
accurate determination of the cut-off, and  $\rm {N_H}$ for the power law, 
which can be seen to determine the cut-off, did not change within the
errors. The lack of change in the LECO is model-independent and is
primary proof that dipping is absent at low energies, since the non-dip
and dip spectra have to converge towards the cut-off. Moreover,
it is clear from Fig. 2 that dipping occurs preferentially at 
higher energies in the PSPC band, and is seen to be due to increased 
absorption of the blackbody. At $\sim $1 keV the non-dip and dip spectra
exhibit the energy-independence well known from {\it Exosat}. The blackbody
contribution (dashed line) shows large changes in $\rm {N_H}$  indicating
that dipping is primarily due to absorption of this component. 
Thus below 0.5 keV, the contribution of the blackbody is very small and
so the extent of dipping is markedly reduced compared with
the 0.5 - 2.0 keV band.

\par Typical spectral fitting results for non-dip data have $\rm {N_H}$
for the power law = $\rm {5.29\pm 0.22\cdot 10^{21}\;H\;atom\;cm^{-2}}$,
normalisation I (at 1 keV) = $\rm {0.58\pm 0.05}$ photons $\rm {cm^{-2}\;s^{-1}}$,
$\rm {N_H}$ for the blackbody = $\rm {5.4^{+5.6}_{-0.2} \cdot 10^{21}}$ 
H atom $\rm {cm^{-2}}$ and I = $\rm {0.12\pm 0.02}$ 
photons $\rm {cm^{-2}\;s^{-1}}$ (90\% confidence errors).
There was very good consistency of the results from different sections of
data with only small variations of parameter values, eg $\rm {I_{PL}}$
varied by about 0.016 photons $\rm {cm^{-2}\;s^{-1}}$.
The corresponding blackbody normalisation was non-zero at the $\rm {8\sigma}$ 
level, showing clearly the necessity for this component in fitting the 
spectra of this source. The additional absorption term for the blackbody 
was in all cases close to zero, showing that that during persistent 
emission both spectral components are subject to the same absorption.
The dip data were fitted by the same model with the normalisations
fixed at non-dip values. Allowing the dips to be modelled by normalisation
changes would be to assume that some change in the emission processes
takes place in the source coincident with the absorption during dipping,
which can be discounted as unphysical. For dip data,
the total $\rm {N_H}$ for the blackbody increased to $\rm
{10.4^{+5.6}_{-1.3}\cdot 10^{21}}$ 
H atom $\rm {cm^{-2}}$, with $\rm {N_H}$ for the power law equal to 
$\rm {5.37\pm 0.22\cdot 10^{21}}$ H atom $\rm {cm^{-2}}$.

\begin{figure}
\epsffile{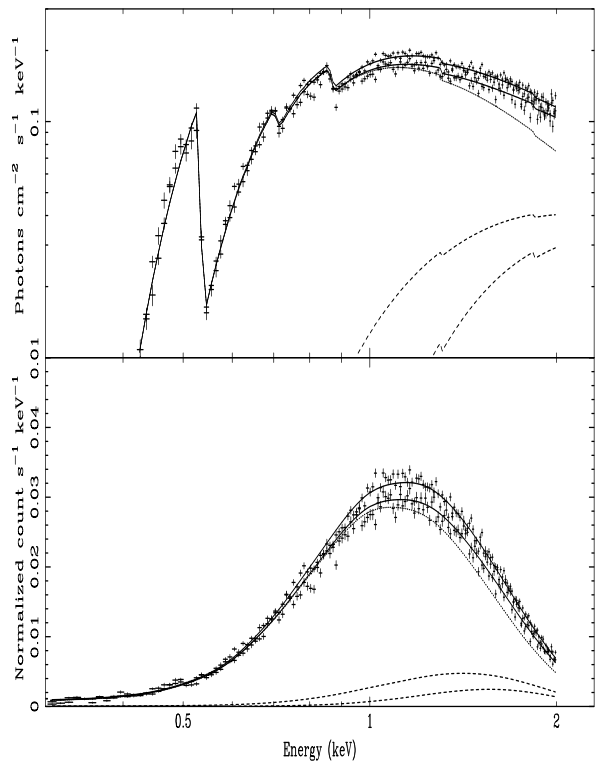}
\caption[]{Spectral fits to non-dip and to dip emission. The solid lines 
show the total fit, and the dotted line and dashed lines show the power law
and blackbody components respectively.}
\end{figure} 

\par Fig. 2
shows that below $\sim $0.5 keV the blackbody is very small and so
implies that below 0.5 keV dipping should be effectively absent 
since in this source, it appears that absorption of the power law in
dipping is very small. To demonstrate this
we have obtained the light curve in the band 0.1 - 0.5 keV shown in Fig.
1b. Although the count rate is of course low in this band, there is
little or no sign of dipping, and this is confirmed by examining the mean count rates
in all of the sections of data in this figure. In the band 0.5 to 2.0 keV, the 
decrease is 3.6 c/s, ie $\rm {10\pm 0.4}$\% which in terms of the standard deviation of the mean 
of non-dip means 0.49 c/s is 7.2$\sigma $, and so is highly significant.
$\rm {N_H}$ for the power law did not change in dipping within the
errors of fitting, and we estimate possible residual dipping in the band
0.1 - 0.5 keV as $\rm {3^{+1}_{-2}}$\% in count rate, as indicated by the bar
in Fig. 1b at the position of dipping (which is 3\% deep).
There may be residual dipping at a low level; however the data of 
Fig. 1b do not constitute a significant detection of this.

\par From the results, absorbed photon fluxes of the spectral components were 
calculated. In the band 0.1 - 2.0 keV, the blackbody comprises 17\% of the
total, whereas in the band 1.0 - 10.0 keV it is 38\% of the total. This is in 
good agreement with a typical blackbody percentage of 39\% taken from our
previous {\it Exosat ME} work. The lower percentage in the PSPC band
corresponds to a blackbody contribution to the count rate of 16\% in the
band 0.1 - 2.0 keV. Thus since the depth of dipping in the {\it Exosat}
observation indicated only partial absorption of the 
blackbody, a level of dipping $\sim $ 10\% in the PSPC band might be 
expected, as is seen.
 
\begin{figure}
\epsffile{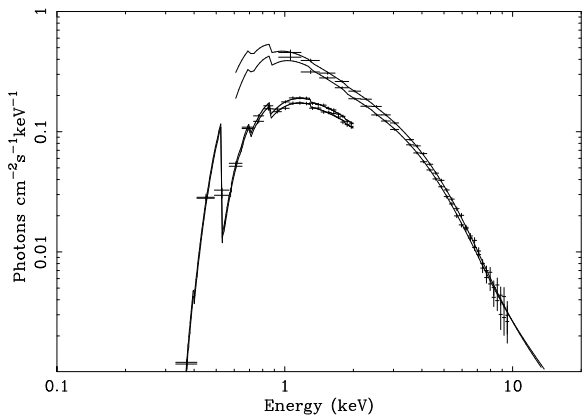}
\caption[]{Comparison of the {\it Rosat PSPC} spectra with typical 
\hbox {{\it Exosat ME}} results for non-dip and dip data.}
\end{figure}

\section{Discussion}

We previously found that a two-component model provides a good explanation
of the energy-independent dipping in the {\it Exosat ME} band for 
X 1755-338, and proposed that this model may be able to explain all of the
dipping sources. We now find that this model gives good fits to the
{\it Rosat PSPC} data and allows the observed features of the PSPC
data to be understood.

\par Firstly there is no significant change in the LECO of the data between
non-dip and dip data. The LECO is determined by the power law component
and the spectral fitting results show there is no significant change in $\rm {N_H}$, 
This lack of change would be quite unexpected in terms of 
photoelectric absorption of a single emission component which should show 
a large change in LECO if the absorber has normal abundances.
Thus two emission regions at least {\it have} to 
be present. The two-component model gives a good explanation of the lack 
of change in the LECO, since the blackbody is responsible for dips but 
does not determine the spectrum below 0.5 keV. 

Secondly, the light curves
show a clear decrease of count rate of 10\% in the band 0.5 - 
2.0 keV; in the lower band 0.1 - 0.5 keV there is little evidence for dipping
but a small residual level of dipping of $\rm { 3^{+1}_{-2}}$\% may persist. 
Thus dipping is substantially reduced especially when compared with the
typical 20\% dipping seen in the {\it Exosat ME} band. Spectral fitting
shows that dipping is caused primarily by absorption of the blackbody.
It is interesting that in this source there appears to be little
absorption of the extended power law component.
Comparison of the {\it Rosat PSPC} results with the previous 
{\it Exosat ME} results shows that the brightness of the source was 
$\sim $ 1.7 times less during the {\it Rosat} observation.

\par The blackbody plus power law model was found to be a good fit to
{\it TTM} data on X\thinspace 1755-338 by Pan et al. (1995). However
they were not able to find a best fit model, and several models fitted
equally well including the above model and a multi-temperature disc plus
power law. The latter model would allow the source to be a black hole
candidate. Thus, the conclusion of Pan et al. that their results strongly
support the black hole nature of the source does not seem to be justified.

\par It is clear that the {\it Rosat PSPC} results support the two-component 
model of the dippers. This model predicts that as dipping is due to absorption
of the blackbody it can not extend indefinitely to energies much less than
$\rm {kT_{bb}}$. This is supported by our results which show
for the first time a marked reduction in the extent of dipping at low energies 
in a LMXB dipping source.

\begin{acknowledgements}
      Analysis was carried out at the University of Birmingham 	using the 
facilities of the PPARC Starlink node. We thank the {\it Rosat} Archive
team at MPE, Garching for providing the data, and the referees for
helpful comments.
\end{acknowledgements}

\end{document}